\documentclass{midl} 

\usepackage{multirow}
\usepackage{ulem}
\usepackage{color}
\usepackage{array}
\usepackage{makecell, multirow, booktabs}
\usepackage{mwe} 
\jmlrvolume{-- Under Review}
\jmlryear{2022}
\jmlrworkshop{Full Paper -- MIDL 2022 submission}
\editors{Under Review for MIDL 2022}

\title[Cross Teaching between CNN and Transformer]{Semi-Supervised Medical Image Segmentation via Cross Teaching between CNN and Transformer}

\midlauthor{\Name{Xiangde Luo\nametag{$^{1,2}$}}\Email{xiangde.luo@std.uestc.edu.cn}\\
\Name{Minhao Hu\nametag{$^{3}$}} \Email{huminhao@sensetime.com}\\
\Name{Tao Song\nametag{$^{3}$}} \Email{songtao@sensetime.com}\\
\Name{Guotai Wang\nametag{$^{1,2,*}$}} \Email{guotai.wang@uestc.edu.cn}\\
\Name{Shaoting Zhang\nametag{$^{1,2,3}$}} \Email{zhangshaoting@uestc.edu.cn}\\
\addr $^{1}$ University of Electronic Science and Technology of China, Chengdu, China \\
\addr $^{2}$ Shanghai AI Lab, Shanghai, China; 
\addr $^{3}$ SenseTime, Shanghai, China;
\addr $^{*}$ Corresponding author}

\begin{document}

\maketitle

\begin{abstract}
Recently, deep learning with Convolutional Neural Networks (CNNs) and Transformers has shown encouraging results in fully supervised medical image segmentation. However, it is still challenging for them to achieve good performance with limited annotations for training. This work presents a very simple yet efficient framework for semi-supervised medical image segmentation by introducing the cross teaching between CNN and Transformer. Specifically, we simplify the classical deep co-training from consistency regularization to cross teaching, where the prediction of a network is used as the pseudo label to supervise the other network directly end-to-end. Considering the difference in learning paradigm between CNN and Transformer, we introduce the Cross Teaching between CNN and Transformer rather than just using CNNs. Experiments on a public benchmark show that our method outperforms eight existing semi-supervised learning methods just with a more straightforward framework. Notably, this work may be the first attempt to combine CNN and transformer for semi-supervised medical image segmentation and achieve promising results on a public benchmark. The code will be released at \url{https://github.com/HiLab-git/SSL4MIS}.
\end{abstract}
\section{Introduction}
Medical image segmentation is a very basic and important step for computer-assisted diagnosis, treatment planning, and intervention~\cite{Wang18,luo2021mideepseg}. Recently, Convolutional Neural Networks (CNNs) and Transformers with large-scale fine annotated images have achieved very promising results, even some applications have been used in the clinical flow~\cite{shi2020clinically,chen2021deep_ro}. But these methods almost require pixel/voxel-level expert labeling, which is more expensive and time-consuming than the natural image annotation~\cite{yu2019uncertainty}. This dilemma makes semi-supervised segmentation a cheap and practical method to train powerful models with limited carefully labeled data and huge unlabeled or roughly labeled data. These proprieties can be used to accelerate clinical data annotation, model development, and even reduce the annotation that is often given by radiologists~\cite{luo2020semi,luo2021urpc,xia20203d,wang2021semi}.\\ \textbf{Semi-supervised medical image segmentation: }Recently, semi-supervised learning has raised high attention in the medical image computing community. {A lot of semi-supervised methods have been proposed for medical image analysis, including pseudo-labelling~\cite{wang2021semi,bai2017semi,chen2021semi}, deep co-training~\cite{qiao2018deep,zhou2019semi}, deep adversarial learning~\cite{zhang2017deep,hu2020coarse}, few-shot learning~\cite{tang2021recurrent}, mean teacher and its extensions~\cite{tarvainen2017mean,yu2019uncertainty,li2020transformation,reiss2021every,you2021momentum,you2021simcvd}, multi-task learning~\cite{luo2020semi,kervadec2019curriculum,chen2019multi}, confidence learning~\cite{vu2019advent}, contrastive learning~\cite{peng2021self}, and etc.} All these methods combine both labeled and unlabeled data to train powerful and robust CNN models. \\
\textbf{CNNs \textit{vs} Transformers for medical image segmentation: }CNN-based medical image segmentation approaches have been studied for many years, and most of them are based on UNet~\cite{ronneberger2015u} or its variants, achieving very promising results in various tasks~\cite{isensee2021nnu}. Although the exceptional representation capacity, CNN-based methods are also limited by lacking the ability of modeling the global and long-range semantic information interaction, due to the intrinsic locality of convolution operations~\cite{chen2021transunet}. More recently, self-attention-based architectures~\cite{dosovitskiy2020image} (vision transformers) are introduced to the vision recognition tasks to model the long-range dependencies. After that, many variants of vision transformers achieved great success in natural image recognition tasks, like Swin-Transformer~\cite{liu2021swin}, DieT~\cite{touvron2021training}, PVT~\cite{wang2021pyramid}, TiT~\cite{han2021transformer}, etc. Benefiting from the great representation capacity of transformers, several works attempt to use transformers to replace or combine CNNs for better medical image segmentation results, such as TransUNet~\cite{chen2021transunet}, Swin-UNet~\cite{cao2021swin}, CoTr~\cite{xie2021cotr}, UNETR~\cite{hatamizadeh2021unetr}, nnFormer~\cite{zhou2021nnformer}, etc. All these works show that transformers can further lead to performance gain than CNNs and also point out that it is worth to pay more attention to the transformer in the future. Although transformers have very exceptional representation capacity, it is still a data-hungry solution for recognition tasks, even require more data than CNNs~\cite{he2021masked,tang2021selfsupervised,you2022class}. How to train transformers with a semi-supervised fashion is also an interesting and challenging problem, especially for data limited medical image analysis tasks.
\par In this work, we present a simple yet efficient regularization scheme between CNN and Transformer, called Cross Teaching between CNN and Transformer. This framework takes both labeled and unlabeled images as inputs, and each input image passes a CNN and a transformer respectively to produce the prediction. For the labeled data, the CNN and transformer are supervised by the ground truth individually. Inspired by~\cite{qiao2018deep,han2018co}, we used predictions of unlabeled images generated by CNN/Transformer to update the parameters of the Transformer/CNN respectively. The advantages of the proposed are two-fold: (1) cross teaching is implicit consistency regularization, which can produce more stable and accurate pseudo labels than explicit consistency regularization. Explicit consistency regularization enforces to minimize the difference of different networks' predictions and optimize them at the same time, it could lead to predictions of different network are same but predictions are wrong. (2) this framework benefits from the two different learning paradigms, CNNs focus on the local information and transformers model the long-range relation, so the cross teaching can help to learn a unified segmenter with these two properties at the same time. {The main \textbf{contributions} are two-fold: (1) We present a simple yet efficient cross teaching scheme for semi-supervised medical image segmentation. The proposed scheme implicitly encourages the consistency between different networks, when the advantages of CNNs and Transformers are leveraged to compensate each other for better performance; (2) To the best of our knowledge, this is the first attempt to use transformers to perform the semi-supervised medical image segmentation task and demonstrate it can outperform eight existing semi-supervised methods on a public benchmark.}

\begin{figure}[t]
\floatconts
  {fig:pipeline}
  {\caption{Overview of \textbf{Cross Teaching between CNN and Transformer}.}}
  {\includegraphics[width=0.81\linewidth]{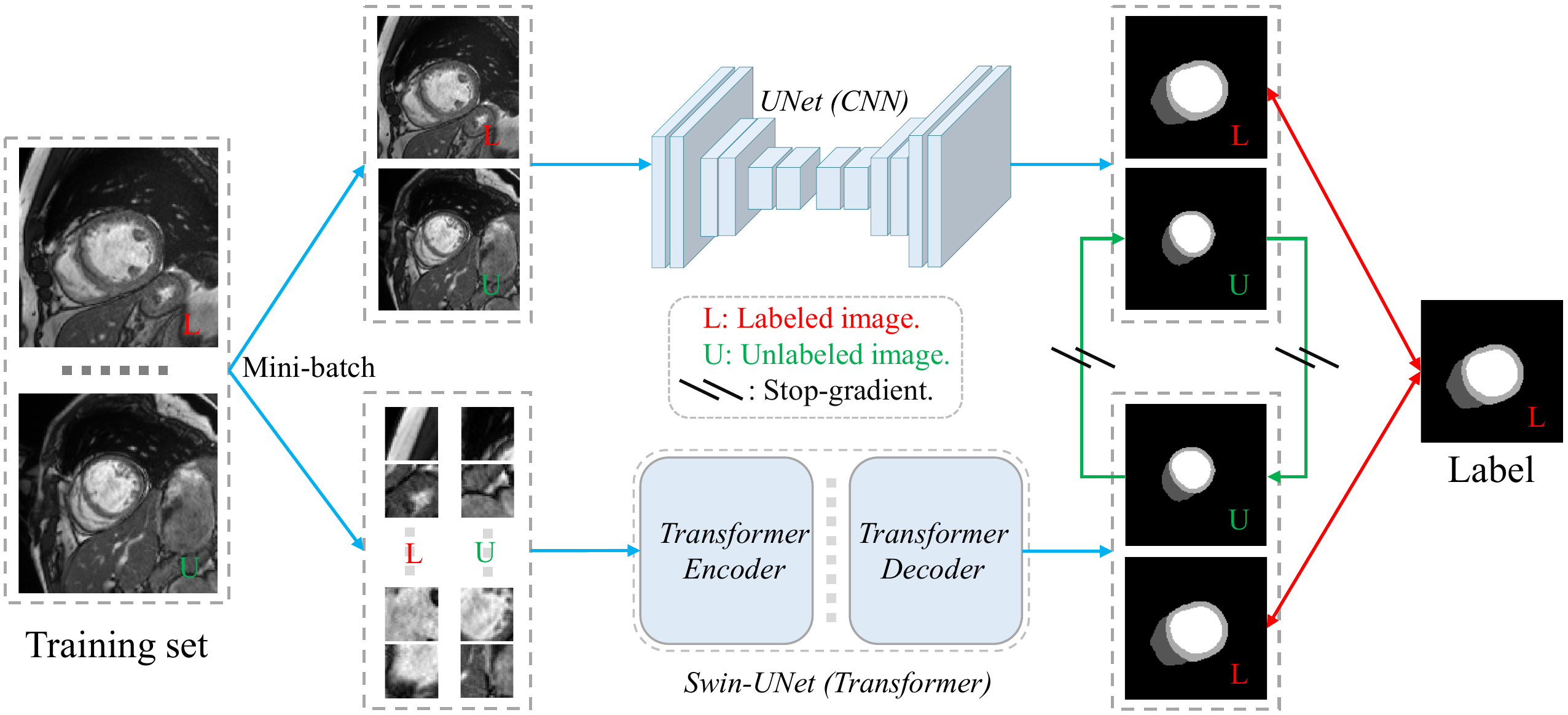}}
\end{figure}
\section{Method}
For the general semi-supervised learning, the training set always consists of two parts: labeled data set ${D}^{l}_{N}$ with $N$ annotated images and unlabeled data set ${D}^{u}_{M}$ with $M$ raw images ($M$ $>>$ $N$), the entire train set is ${D}_{N+M} = {D}^{l}_{N} \cup {D}^{u}_{M}$. For an image $x_i$ $\in$ ${D}^{l}$, its ground truth $y_i$ is available. In contrast, if $x_i$ $\in$ ${D}^{u}$, its ground truth is not provided. In this work, the proposed Cross Teaching between CNN and Transformer is depicted in Fig.~\ref{fig:pipeline}. If $x_i$ $\in$ ${D}^{l}$, a commonly-used supervised loss function is used to update models' parameters. When $x_i$ belongs to ${D}^{u}$, we use a cross teaching strategy to cross supervise between a CNN ($f^c_\phi(.)$) and a Transformer ($f^t_\phi(.)$) for the updating of the parameters.

\subsection{Cross teaching between CNN and Transformer}
The original idea of cross teaching is inspired by three existing works: Deep Co-Training~\cite{qiao2018deep}, Co-Teaching~\cite{han2018co} and Cross Pseudo Supervision~\cite{chen2021semi}. Deep Co-Training~\cite{qiao2018deep} trains multiple deep neural networks with different views inputs and encourages view consistency for semi-supervised learning. Co-Teaching trains two deep neural networks simultaneously, and lets them teach each other in a mini-batch for noise-robust learning. Cross Pseudo Supervision~\cite{chen2021semi} trains two networks with the same architecture and different initializations to teach each other in a mini-batch for semi-supervised learning. All these methods introduce perturbations and encourage prediction to be consistent during the training stage. The differences are that Deep Co-Training uses input-level perturbation (multi-views), Co-Teaching utilizes the supervision-level perturbation (noisy labels), and Cross Pseudo Supervision introduces perturbation in the network architecture-level. Here, we introduce the perturbation in both learning paradigm-level and output-level. For an input image $x_i$, the proposed framework produces two predictions:
\begin{equation}
    p^c_i = f^c_\phi(x_i);~~~p^t_i = f^t_\phi(x_i)
\end{equation} where $p^c_i$, $p^t_i$ represent the prediction of a CNN ($f^c_\phi(.)$) and a Transformer ($f^t_\phi(.)$), respectively. As previously mentioned, CNN and Transformer are different learning paradigms for vision recognition, where CNN relies on the local convolution operation and the Transformer is based on the long-range self-attention, so these predictions have different properties essentially in the output level. Based on the predictions of $f^c_\phi(.)$ and $f^t_\phi(.)$, the pseudo labels for the cross teaching strategy are generated by this way:
\begin{equation}
    pl^c_i = argmax(p^t_i) = argmax(f^t_\phi(x_i));~~~pl^t_i = argmax(p^c_i) = argmax(f^c_\phi(x_i))
\end{equation} where $pl^c_i$, $pl^t_i$ are generated pseudo labels for the CNN ($f^c_\phi(.)$) and the Transformer ($f^t_\phi(.)$) training, respectively. It's worthy to point that $pl^c_i$, $pl^t_i$ are pseudo segmentation results, and there is no gradient back-propagation between $p^c_i$ and $pl^c_i$, and between $p^t_i$ and $pl^t_i$ in each mini-batch. Then, the cross teaching loss for the unlabeled data is defined as:
\begin{equation}
    \mathcal{L}_{ctl} = \underbrace{\mathcal{L}_{dice}(p^c_i, pl^c_i)}_{supervision~for~CNNs} + \underbrace{\mathcal{L}_{dice}(p^t_i, pl^t_i)}_{supervision~for~Transformers}
\end{equation}where $\mathcal{L}_{dice}$ is the standard dice loss function. Differently from consistency regularization loss, the cross teaching loss is a bidirectional loss function, one stream is from the CNN to the Transformer and the other is the Transformer to the CNN, there are no explicit constraints to enforce their predictions to become similar. In our framework, the transformer is also just used for complementary training, not used to produce final predictions.
\subsection{The overall objective function}
The overall training objective function is a joint loss with two parts, a supervised loss on the labeled data and an unsupervised loss for the unlabeled data. The supervised loss $\mathcal{L}_{sup}$ consists of two widely-used loss functions:
\begin{equation}
    \mathcal{L}_{sup} = \mathcal{L}_{ce}(p_i, y_i) + \mathcal{L}_{dice}(p_i, y_i)
\end{equation}where $\mathcal{L}_{ce}$, $\mathcal{L}_{dice}$ are the cross-entropy loss and dice loss, respectively. $p_i$, $y_i$ represent the prediction and label of image $x_i$. The overall objective is defined as :
\begin{equation}
    \mathcal{L}_{total} = \mathcal{L}_{sup} + \lambda\mathcal{L}_{ctl}
\end{equation}where $\lambda$ is a weight factor, which is defined by a time-dependent Gaussian warming up function commonly~\cite{yu2019uncertainty,luo2020semi,ssl4mis2020}: $\lambda(t) = 0.1 \cdot e^{(-5(1-\frac{t_i}{t_{total}})^2)}$, where $t_i$ denotes the current training iteration and $t_{total}$ is the total iteration number.
\section{Experiments}
\subsection{Dataset and evaluation metrics}
In this work, all experiments and comparisons are based on the public benchmark dataset ACDC~\cite{bernard2018deep}. The ACDC dataset contains 200 annotated short-axis cardiac cine-MR images from 100 patients. The segmentation masks of the left ventricle (LV), myocardium (Myo), and right ventricle (RV) are provided for clinical and algorithm research. 140 images from 70 patients, 60 images from 30 patients are randomly selected for training and validation respectively. Due to the large inter-slice spacing, 2D segmentation is more suitable than direct 3D segmentation~\cite{bai2017semi}. For the pre-processing, we resize all the slices into 256$\times$256 pixels and re-scale the intensity of each slice to [0, 1]. We used standard data augmentation to enlarge training sets' scale and avoid over-fitting, including random cropping with a 224$\times$224 patch, random rotating between -25 and 25 degrees, random flipping. At the inference stage,  predictions are generated slice by slice and stacked into a 3D volume. For a fair comparison, we don't use any post-processing strategy. Two commonly-used metrics are employed to quantitatively evaluate the 3D segmentation results: 1) Dice Coefficient (DSC); 2) 95\% Hausdorff Distance (HD$_{95}$). \textcolor{red}{(Details in Sec.\ref{sec:dataset_details}).}
\subsection{Implementation details}
\textbf{Network architectures and training details: }There are two types of networks in the proposed method (see Fig.~\ref{fig:pipeline}): a CNN-based segmentation network UNet~\cite{ronneberger2015u} and a Transformer segmentation Swin-UNet~\cite{cao2021swin}. Both of them are U-shape-based networks, but they are based on two different learning paradigms. For a fair comparison, in this work, we employ the open-sourced implementation of UNet~\cite{ronneberger2015u} and Swin-UNet~\cite{cao2021swin} as baselines. We use PyTorch~\cite{paszke2019pytorch} for all method's implementations, and run all experiments on a Ubuntu desktop with a GTX1080TI GPU. All these networks are trained by the SGD optimizer with a batch size of 16, where half of them are labeled in batch for semi-supervised learning. The poly learning rate strategy is used to adjust the learning rate, where the initial learning rate is set to 0.01. All implementation are available at: \url{https://github.com/HiLab-git/SSL4MIS}.\\
\textbf{Ablation study: }In this work, we first investigate the performance when using the Transformers for semi-supervised learning directly. We use the Swin-UNet~\cite{cao2021swin} as backbones to replace the UNet~\cite{ronneberger2015u} for semi-supervised learning in Mean Teacher (MT)~\cite{tarvainen2017mean}, Entropy Minimization (Ent Min)~\cite{vu2019advent}, Deep Adversarial Network (DAN)~\cite{zhang2017deep}, Deep Co-Training (DCT)~\cite{qiao2018deep}. Then, we further investigate the results when cross teaching between different architectures, including CNN and Transformer, CNNs and CNN and Transformer and Transformers. Finally, we investigate the performance gain of different cross teaching loss functions, including cross-entropy loss, dice loss, and compare them against classical consistency regularization~\cite{tarvainen2017mean,yu2019uncertainty}.\\
\textbf{Comparison with baselines and existing methods: }{To demonstrate the effectiveness of the proposed method, we compared our method against baselines and eight recently semi-supervised methods. Firstly, we investigate the upper-bound/low-bound performances of UNet and Swin-UNet based on all/limited labeled images, respectively, referred to as full/limited supervisions (FS/LS).} Then, we compared with eight previous semi-supervised methods: 1) Mean Teacher (MT)~\cite{tarvainen2017mean}, 2) Entropy Minimization (EM)~\cite{vu2019advent}, 3) Deep Adversarial Network (DAN)~\cite{zhang2017deep}, 4) Uncertainty Aware Mean Teacher (UAMT)~\cite{yu2019uncertainty}, 5) Interpolation Consistency Training (ICT)~\cite{ijcai2019-504}, 6) Cross Pseudo Supervision (CPS)~\cite{chen2021semi}, 7) Cross Consistency Training (CCT)~\cite{ouali2020semi}, 8) Deep Co-Training (DCT)~\cite{qiao2018deep}. All these methods used the same backbone and were trained and tested with the same settings, notably all these methods are open available~\cite{ssl4mis2020}.
\section{Results}

\begin{table}[t]
\centering
\tiny\setlength{\extrarowheight}{1pt}
\renewcommand\arraystretch{0.05}
\caption{Ablation study results when using 7 cases as labeled. RV, Myo, LV represent the right ventricle, myocardium and left ventricle, respectively. The first section shows the results when using transformers for semi-supervised segmentation directly. The second section lists the results when using different architectures and supervision strategies, where CT and CR mean the cross teaching and consistency regularization, respectively. The last section shows the effects when using different cross-teaching loss functions. $^*$ means the predictions are produced by Swin-UNet, the others are based on UNet. Blue numbers mean the best results.}
\setlength{\tabcolsep}{0.02mm}{
\begin{tabular}{lllllllll}
\hline
\multirow{2}{*}{Method}&\multicolumn{2}{c}{\textbf{RV}}&\multicolumn{2}{c}{\textbf{Myo}}&\multicolumn{2}{c}{\textbf{LV}}&\multicolumn{2}{c}{\textbf{Mean}} \\\cline{2-9} 
&\textit{DSC}&\textit{$HD_{95}$}&\textit{DSC}&\textit{$HD_{95}$}&\textit{DSC}&\textit{$HD_{95}$}&\textit{DSC}&\textit{$HD_{95}$}\\
\hline
LS$^*$&0.42(0.21)&34.6(24.6)&0.499(0.18)&19.0(12.2)&0.617(0.23)&23.6(13.4)&0.512(0.207)&25.7(16.8) \\
MT$^*$&0.433(0.231)&25.3(19.6)&0.486(0.179)&18.2(11.9)&0.614(0.234)&22.0(13.0)&0.511(0.215)&21.8(14.8) \\
DAN$^*$&0.504(0.208)&24.0(18.0)&0.438(0.159)&18.5(14.2)&0.643(0.2)&32.3(27.6)&0.528(0.189)&24.9(19.9) \\
DCT$^*$&0.465(0.223)&29.0(22.2)&0.499(0.175)&17.5(12.1)&0.622(0.22)&22.5(13.0)&0.528(0.206)&23.0(15.8) \\
EM$^*$&0.456(0.216)&32.1(24.2)&0.51(0.176)&18.9(11.7)&0.62(0.223)&23.9(12.3)&0.529(0.205)&25.0(16.1) \\
FS$^*$&0.785(0.137)&11.4(13.7)&0.779(0.083)&5.6(7.4)&0.863(0.123)&7.4(9.8)&0.809(0.115)&8.1(10.3) \\
\hline
CNN\&CNN(CT)&0.791(0.189)&13.2(15.7)&0.821(0.067)&8.4(11.6)&0.886(0.092)&11.5(16.2)&0.833(0.116)&11.0(14.5)\\
Trans\&Trans(CT)$^*$&0.805( 0.117)&12.9(18.0)&0.779(0.072)&\textcolor{blue}{6.1(7.2)}&0.856(0.119)&12.2(15.3)&0.813(0.103)&10.4(13.5)\\
CNN\&Trans(CR)&0.782(0.195)&14.4(15.3)&0.806(0.082)&12.7(15.5)&0.87(0.111)&18.1(21.3)&0.82(0.129)&15.1(17.4)\\
CNN\&Trans(\textbf{proposed})&\textcolor{blue}{0.848(0.112)}&\textcolor{blue}{7.80(7.6)}&\textcolor{blue}{0.844(0.052)}&6.9(9.2)&\textcolor{blue}{0.901(0.085)}&\textcolor{blue}{11.2(14.8)}&\textcolor{blue}{0.864(0.083)}&\textcolor{blue}{8.6(10.5)}\\
\hline
Ours(CE)&0.807(0.183)&9.8(11.0)&0.829(0.066)&7.1(9.1)&0.881(0.106)&14.4(17.8)&0.839(0.118)&10.5(12.6)\\
Ours(CE)$^*$&0.829(0.103)&10.6(16.0)&0.802(0.065)&\textcolor{blue}{5.1(5.6)}&0.875(0.109)&11.5(17.4)&0.835(0.092)&9.1(13.0)\\
Ours(DICE)&0.848(0.112)&\textcolor{blue}{7.80(7.6)}&0.844(0.052)&6.9(9.2)&\textcolor{blue}{0.901(0.085)}&11.2(14.8)&\textcolor{blue}{0.864(0.083)}&8.6(10.5)\\
Ours(DICE)$^*$&\textcolor{blue}{0.85(0.099)}&9.2(15.2)&0.825(0.056)&5.4(7.2)&0.893(0.09)&\textcolor{blue}{8.4(12.9)}&0.856(0.082)&7.7(11.8)\\
Ours(CE+DICE)&0.84(0.142)&8.8(9.6)&\textcolor{blue}{0.849(0.045)}&6.0(7.7)&0.9(0.089)&12.4(18.9)&0.863(0.092)&9.1(12.1)\\
Ours(CE+DICE)$^*$&0.844(0.098)&8.1(11.6)&0.815(0.06)&5.6(7.8)&0.886(0.097)&9.1(13.1)&0.848(0.085)&\textcolor{blue}{7.6(10.8)}\\
\hline
\end{tabular}}
\label{tab:abl_vit}
\end{table}

\begin{table}[t]
\centering
\tiny
\renewcommand\arraystretch{0.25}
\caption{Mean 3D \textit{DSC} and \textit{$HD_{95}$} (mm) on the ACDC dataset. All results are based on the same backbone (UNet) with a fixed seed. Mean and standard variance (in parentheses) are presented in this table. Red numbers denote the p-value $<$ 0.05 based on paired t-test when comparing with existing methods. The performance of nnUNet is borrowed from ACDC leaderboard.}
\setlength{\tabcolsep}{0.25mm}{
\begin{tabular}{llllllllll}
\hline
\multirow{2}{*}{Labeled}&\multirow{2}{*}{Method}&\multicolumn{2}{c}{\textbf{RV}}&\multicolumn{2}{c}{\textbf{Myo}}&\multicolumn{2}{c}{\textbf{LV}}&\multicolumn{2}{c}{\textbf{Mean}} \\\cline{3-10} 
&&\textit{DSC}&\textit{$HD_{95}$}&\textit{DSC}&\textit{$HD_{95}$}&\textit{DSC}&\textit{$HD_{95}$}&\textit{DSC}&\textit{$HD_{95}$}\\
\hline
&LS&0.37(0.32)&
44.4(28.4)&
0.548(0.287)&
24.4(19.9)&
0.618(0.329)&
24.3(21.4)&
0.512(0.312)&
31.0(23.2)\\
&MT&0.403(0.291)&
53.9(28.7)&
0.586(0.251)&
23.1(20.5)&
0.709(0.246)&
26.3(25.8)&
0.566(0.263)&
34.5(25.0)\\
&DAN&0.378(0.318)&
39.6(25.7)&
0.568(0.259)&
25.8(20.3)&
0.64(0.306)&
32.4(27.9)&
0.528(0.294)&
32.6(24.6)\\
\multirow{10}{*}{3 cases}&DCT&0.413(0.305)&
31.7(20.2)&
0.617(0.235)&
20.3(18.0)&
0.717(0.254)&
27.3(24.5)&
0.582(0.265)&
26.4(20.9)\\
&EM&0.447(0.312)&
32.4(22.7)&
0.628(0.235)&
19.0(20.3)&
0.731(0.244)&
20.9(20.2)&
0.602(0.264)&
24.1(21.0)\\
&UAMT&0.508(0.328)&
35.4(24.8)&
0.615(0.26)&
19.3(21.8)&
0.707(0.273)&
22.6(19.8)&
0.610(0.287)&
25.8(22.1)\\
&ICT&0.448(0.327)&
23.8(16.4)&
0.620(0.24)&
20.4(20.4)&
0.673(0.286)&
24.1(21.1)&
0.581(0.284)&
22.8(19.3)\\
&CCT&0.408(0.322)&
34.2(23.6)&
0.647(0.206)&
22.4(19.4)&
0.704(0.239)&
27.1(23.9)&
0.586(0.256)&
27.9(22.3)\\&CPS&0.438(0.306)&
35.8(24.1)&
0.652(0.213)&
18.3(16.3)&
0.72(0.248)&
22.2(22.0)&
0.603(0.256)&
25.5(20.8)\\
&\textbf{Ours}&\textcolor{red}{0.577(0.338)}&
21.4(20.2)&
0.628(0.251)&
\textcolor{red}{11.5(11.7)}&
\textcolor{red}{0.763(0.244)}&
\textcolor{red}{15.7(15.5)}&
\textcolor{red}{0.656(0.278)}&
\textcolor{red}{16.2(15.8)}\\
\hline
&LS&0.649(0.31)&
19.4(19.5)&
0.787(0.087)&
12.2(13.8)&
0.856(0.117)&
17.4(18.6)&
0.764(0.171)&
16.3(17.3)\\
&MT&0.769(0.198)&
17.9(21.3)&
0.794(0.093)&
10.8(13.8)&
0.866(0.105)&
14.5(17.6)&
0.810(0.132)&
14.4(17.6)\\
&DAN&0.763(0.212)&
13.5(15.1)&
0.784(0.105)&
9.8(10.0)&
0.840(0.138)&
20.6(21.0)&
0.795(0.152)&
14.6(15.4)\\
&DCT&0.750(0.221)&
15.3(15.6)&
0.793(0.098)&
10.7(12.6)&
0.870(0.099)&
15.5(19.2)&
0.804(0.14)&
13.8(15.8)\\
\multirow{10}{*}{7 cases}&EM&0.729(0.238)&
15.3(14.7)&
0.79(0.099)&
12.4(16.2)&
0.855(0.122)&
15.8(17.4)&
0.791(0.153)&
14.5(16.1)\\
&UAMT&0.775(0.202)&
11.5(10.9)&
0.801(0.092)&
13.7(24.9)&
0.871(0.103)&
18.1(20.9)&
0.815(0.132)&
14.4(18.9)\\
&ICT&0.755(0.232)&
11.4(11.5)&
0.807(0.082)&
8.7(9.2)&
0.871(0.101)&
14.0(17.3)&
0.811(0.138)&
11.4(12.7)\\
&CCT&0.766(0.198)&
14.3(16.0)&
0.812(0.071)&
10.4(13.4)&
0.87(0.111)&
14.6(18.6)&
0.816(0.127)&
13.1(16.0)\\
&CPS&0.791(0.189)&
13.2(15.7)&
0.821(0.067)&
8.4(11.6)&
0.886(0.092)&
11.5(16.2)&
0.833(0.116)&
11.0(14.5)\\
&\textbf{Ours}&\textcolor{red}{0.848(0.112)}&
\textcolor{red}{7.80(7.6)}&
\textcolor{red}{0.844(0.052)}&
\textcolor{red}{6.9(9.2)}&
0.901(0.085)&
11.2(14.8)&
\textcolor{red}{0.864(0.083)}&
\textcolor{red}{8.60(10.5)}\\
\hline
Total&FS&0.900(0.075)&
4.40(4.9)&
0.893(0.028)&
2.4(4.0)&
0.941(0.048)&
4.0(9.7)&
0.911(0.05)&
3.60(6.2)\\
&nnUNet&0.925&10.1&0.908&7.4&0.948&6.2&0.927&7.9\\
\hline
\end{tabular}}
\label{tab:sota_acdc}
\end{table}
\textbf{Ablation study: }Tab.~\ref{tab:abl_vit} shows the results of ablation study. The first section presents the results of using Swin-UNet as segmentation networks to perform semi-supervised segmentation rather than UNet. It can be found that compared with UNet, Swin-UNet performs quite badly when combing previous semi-supervised approaches (results of UNet are presented in Tab.~\ref{tab:sota_acdc}). The reason may be transformers are data-hungry approaches, directly transferring them to perform semi-supervised learning may be not suitable. The second section lists the comparison results when using cross teaching to supervise different combinations. It can be observed that using cross teaching between CNN and Transformer can achieve better results than others combinations. It demonstrates that CNN and Transformer with different learning paradigm can compensate each other in the training stage. In addition, the results also demonstrated that cross teaching strategy outperforms the consistency regularization. Finally, we investigate the impact of different cross-teaching loss functions. The last section shows that dice loss can further improve the performance than cross-entropy loss, but the joint loss of cross-entropy and dice can not lead to more gain. {In addition, we also investigated the performance of the Transformer branch and the ensemble of the Transformer and CNN branches. We found that the Transformer branch achieves very similar results compared with the CNN branch, and the ensemble result outperforms both the Transformer and CNN branches. Still, the transformer and ensemble results require more computation cost, as the transformer has more parameters (27.12$M$ $vs$ 1.81$M$). So, we use the CNN branch outputs to compare with existing methods fairly. (Details in Sec.~\ref{sec:training_details},~\ref{sec:results_ana} and~\ref{sec:performance_gap})}\\
\textbf{Comparison with baselines and existing works: }Tab.~\ref{tab:sota_acdc} lists the results of all methods on the ACDC dataset when using 3 cases (6 volumes) and 7 cases (14 volumes) as labeled samples. Note that, all differences between these methods just exist in the training stage. In the inference stage, these methods just used the trained UNet to produce final predictions and ignored all auxiliary training modules, and also did not use any ensemble strategies. It can be found that the proposed cross-training between CNN and Transformer achieves better results than the other 8 methods in a large margin when used 3 and 7 labeled cases. In the setting of 7 labeled cases, the proposed achieved a significant gain than the second method (cross pseudo supervision strategy), in terms of 3.8\% in $DSC$ and 3.6$mm$ in $HD_{95}$ respectively. Furthermore, despite the labeled sample being very limited (3 cases are labeled, which is less than 10\% of the training set), our proposed method remains is superior to other methods, with more than 4\% of mean $DSC$ improvement and 6$mm$ of mean $HD_{95}$ decrease. These results show the potential of the proposed method which can alleviate the label cost by learning from the limited data and large-scale unlabeled data. In addition, compared with existing semi-supervised learning methods, the proposed framework consists of very simple training strategies and common components with very low complexity, it is more desirable to apply it in clinical practice. {Although the proposed outperforms existing semi-supervised methods, it also can't achieve a comparable results compared with the state-of-the-art (SOTA) fully-supervised method~\cite{isensee2021nnu}. It shows that using semi-supervised methods to achieve SOTA remains an important yet challenging problem.} Fig.~\ref{fig:vis} shows some visualization comparisons between various methods when using 3 labeled cases and 7 labeled cases. Compared with CCT~\cite{ouali2020semi} and CPS~\cite{chen2021semi}, our method can produce more plausible segmentation with fewer false-positive and missing-segmentation regions.
\begin{figure}[t]
\floatconts
  {fig:vis}
  {\caption{Visualization comparison of different methods on validation images. The first two rows used 3 labeled cases and the last two rows used 7 labeled cases.}}
  {\includegraphics[width=0.73\linewidth]{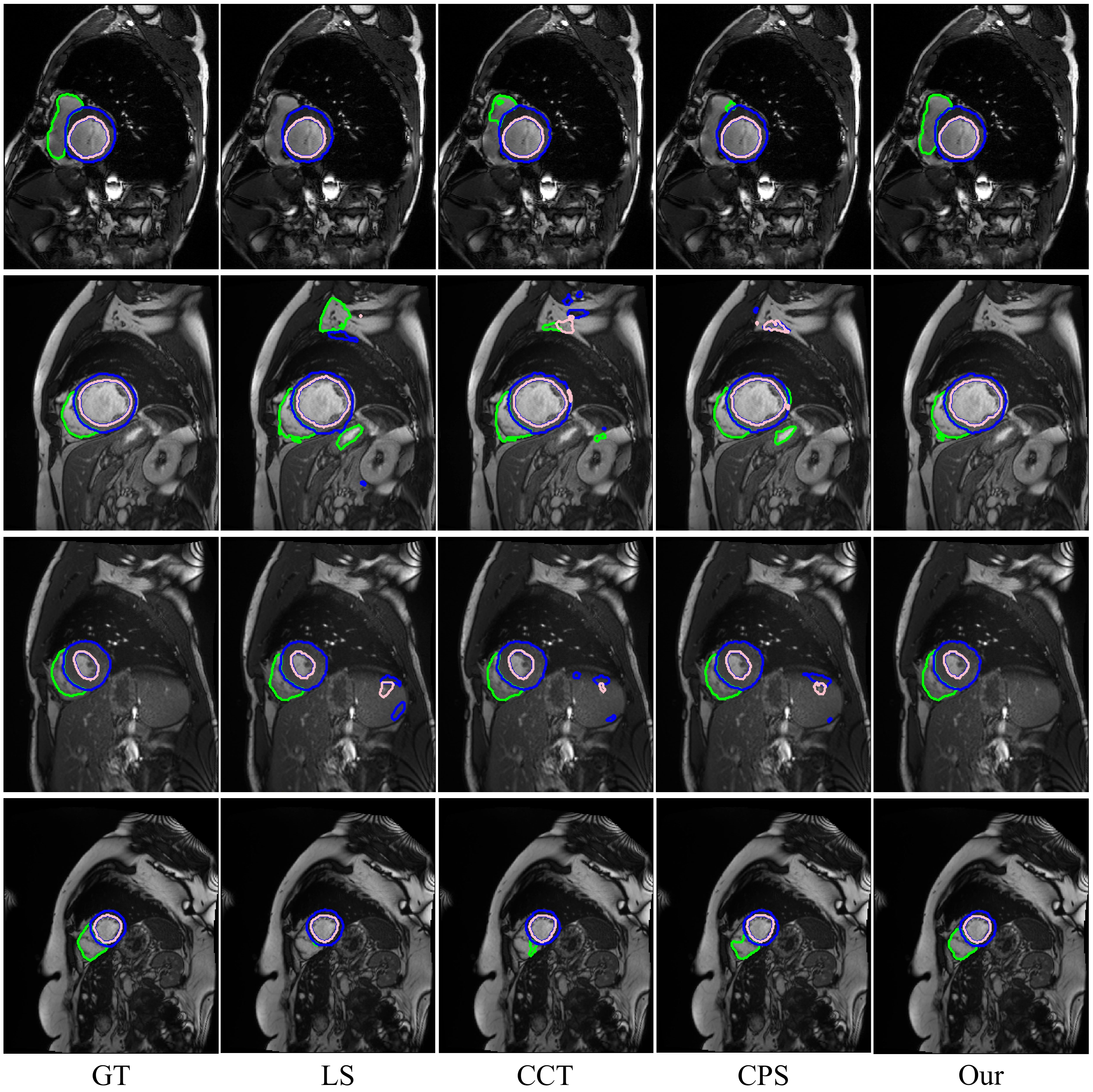}}
\end{figure}

\section{Conclusion}
{In conclusion, this work introduces the Transformer to the semi-supervised medical image segmentation.} To achieve this goal, inspired by co-teaching~\cite{han2018co} and cross pseudo supervision~\cite{chen2021semi}, we present a cross teaching between CNN and Transformer to utilize the unlabeled data. The idea is based on the assumption that CNN can capture local features efficiently and Transformer can model the long-range relation better, and these properties can complement each other during training. Experimental results on an open-benchmark showed that the proposed can outperform eight existing semi-supervised learning methods. In the future, we will combine some advanced techniques~\cite{tang2021selfsupervised} with this method to further reduce the annotation costs, especially for the dense-annotation-based multi-organ segmentation tasks~\cite{luo2021word}.

\bibliography{ref}

\appendix

\section{{Details of experiments}}
\subsection{Dataset}\label{sec:dataset_details}The ACDC dataset\footnote{https://www.creatis.insa-lyon.fr/Challenge/acdc/index.html} consists of 200 short-axis cine-MRI scans from 100 patients, and each patient has two annotated scans corresponding to end-diastolic (ED) and end-systolic (ES) phases. In this paper, we split the dataset based on patient-id into the training set (70 patients) and validation set (30 patients), so the training set has 140 scans (70 ED scans and 70 ES scans), and the validation set consists of 60 scans (30 ED scans and 30 ES scans). For semi-supervised learning, we randomly select 3 patients (6 scans) and 7 patients (14 scans) as labeled data ($\approx$ 5\% and 10\% of the training set) and the remaining 67 patients (144 scans) and 63 patients (126 scans) are seen as unlabeled data.
\subsection{Training details}\label{sec:training_details}
We used UNet~\cite{ronneberger2015u} and Swin-UNet~\cite{cao2021swin} as segmentation networks respectively, their PyTorch implementations borrowed from widely-used projects PyMIC\footnote{https://github.com/HiLab-git/PyMIC} and Swin-UNet\footnote{https://github.com/HuCaoFighting/Swin-Unet}. All experiments of this paper were run on a Ubuntu 16.04 desktop with a GTX1080TI GPU and the PyTorch1.8.1 library. All models are optimized using stochastic gradient descent (SGD) with the polylearning rate strategy, where the initial learning rate was set to 0.01. The total iterations are 30k, and we used the latest checkpoint for testing and reporting results~\cite{yu2019uncertainty}. For a fair comparison, all existing semi-supervised segmentation methods used UNet~\cite{ronneberger2015u} as the segmentation network and also used the same training strategies (data augmentation methods, total iterations, and other training hyper-parameters). To reproduce these results, all methods' training and testing code and processed data are available at: \url{https://github.com/HiLab-git/SSL4MIS}. 
\section{{Results analysis}}
\subsection{Computational-cost}\label{sec:results_ana}
To compare the computational cost among the proposed and others comparisons, we investigated the number of forwarding pass times of an input image in one iteration (\textit{FTimes}), the total training time (\textit{TTimes}), per case inference time (\textit{ITimes}), when using same software and hardware settings. Table~\ref{tab:comput-cost} lists the quantitative comparison of computational-cost. Note that all existing methods used UNet as a segmentation network, and our method used UNet and Swin-UNet at the same time, so we listed UNet/Swin-UNet/ensemble of UNet and Swin-UNet inference time, respectively. It can be found that our method requires more training time than others, as the Swin-UNet consists of more parameters. But all methods require very similar inference costs when using the same segmentation network. For our method, the transformer branch needs more inference time than the CNN branch due to the transformer with more parameters. The ensemble results spend the most inference time than other methods, as it runs two models to generate predictions simultaneously.

\begin{table*}[hpbt]
\centering
\normalsize
\caption{Comparison of computational cost between our method and existing methods based on 7 labeled cases. \textit{FTimes} means the times of an input image passed the networks during one iteration. \textit{TTime} means the total training time (hours). \textit{ITime} means per case inference time (s). For our method, we listed the inference time of CNN (pink number), Transformer (lime number), and their ensemble (yellow number), respectively.}
\scalebox{0.70}{
\begin{tabular}{lcccccccccc
}
\hline
 &LS&MT&DAN&DCT&EM&UAMT&ICT&CCT&CPS&Ours\\
\hline
\textit{FTimes}&1&2&2&2&1&9&3&1&2&2\\
\textit{TTime(h)}&4.17&4.37&5.33&4.83&4.50&5.61&5.15&5.08&5.17&6.22\\
\textit{ITime(s)}&0.46&0.46&0.46&0.46&0.46&0.46&0.46&0.46&0.46&\textcolor{pink}{0.46}/\textcolor{lime}{0.57}/\textcolor{yellow}{0.78}\\
\hline
\end{tabular}}
\label{tab:comput-cost}
\end{table*}

\subsection{Performance of CNN branch and Transformer branch}\label{sec:performance_gap}
We also investigated the difference of performance in the CNN branch and Transformer branch in 3 and 7 labeled cases settings. The results are presented in Table~\ref{tab:cnn_trans}. It can be found that the CNN branch and Transformer branch achieve very similar performance, but as Sec.\ref{sec:results_ana} shows, the Transformer branch requires more inference cost. We further investigated the performance when ensembling the prediction of CNN and Transformer, the ensemble prediction is defined as $argmax(\frac{f^t_\phi(x_i) + f^c_\phi(x_i)}{2})$. The result was listed in the lost row of Table~\ref{tab:cnn_trans}

\begin{table}
\centering
\tiny
\renewcommand\arraystretch{0.25}
\caption{Mean 3D \textit{DSC} and \textit{$HD_{95}$} (mm) on the ACDC dataset. These results are based on our proposed method, Ours (CNN), Ours (Trans), and Ours CNN \& Trans) represent the result of the CNN branch and the Transformer branch and their ensemble, respectively. Mean and standard variance (in parentheses) are presented in this table.}
\setlength{\tabcolsep}{0.25mm}{
\begin{tabular}{llllllllll}
\hline
\multirow{2}{*}{Labeled}&\multirow{2}{*}{Method}&\multicolumn{2}{c}{\textbf{RV}}&\multicolumn{2}{c}{\textbf{Myo}}&\multicolumn{2}{c}{\textbf{LV}}&\multicolumn{2}{c}{\textbf{Mean}} \\\cline{3-10} 
&&\textit{DSC}&\textit{$HD_{95}$}&\textit{DSC}&\textit{$HD_{95}$}&\textit{DSC}&\textit{$HD_{95}$}&\textit{DSC}&\textit{$HD_{95}$}\\
\hline
\multirow{3}{*}{3 cases}&\textbf{Ours(CNN)}&0.577(0.338)&
21.4(20.2)&
0.628(0.251)&
11.5(11.7)&
0.763(0.244)&
15.7(15.5)&
0.656(0.278)&
16.2(15.8)\\
&\textbf{Ours(Trans)}&\textbf{0.660(0.198)}&
23.1(20.7)&
0.597(0.201)&
10.9(9.6)&
0.748(0.207)&
12.4(10.8)&
0.669(0.202)&
15.5(13.7)\\
&\textbf{Ours(Trans \& CNN)}&0.648(0.283)&\textbf{16.5(12.7)}&
\textbf{0.638(0.244)}&
\textbf{10.5(11.1)}&
\textbf{0.777(0.236)}&
\textbf{11.6(13.5)}&
\textbf{0.688(0.254)}&
\textbf{12.8(12.4)}\\
\hline

\multirow{2}{*}{7 cases}&\textbf{Ours(CNN)}&0.848(0.112)&
7.80(7.6)&
0.844(0.052)&
6.9(9.2)&
0.901(0.085)&
11.2(14.8)&
0.864(0.083)&
8.60(10.5)\\
&\textbf{Ours(Trans)}&0.850(0.099)&
9.20(15.2)&
0.825(0.056)&
5.4(7.2)&
0.893(0.09)&
8.40(12.9)&
0.856(0.082)&
7.70(11.8)\\
&\textbf{Ours(Trans \& CNN)}&\textbf{0.865(0.094)}&
\textbf{6.4(6.0)}&
\textbf{0.851(0.049)}&
\textbf{4.7(6.2)}&
\textbf{0.909(0.078)}&
\textbf{7.5(12.8)}&
\textbf{0.875(0.074)}&
\textbf{6.2(8.3)}\\
\hline
\end{tabular}}
\label{tab:cnn_trans}
\end{table}



\end{document}